\documentclass[12pt,a4paper]{article}
\usepackage[cp850]{inputenc}
\usepackage{amssymb}
\usepackage{latexsym}
\usepackage[dvips]{epsfig}
\usepackage{pstricks,pst-node}
\renewcommand{\baselinestretch}{1.5}
\begin{document}
\begin{titlepage} 
\renewcommand{\baselinestretch}{1}
\small\normalsize
\begin{flushright}
hep-th/0508202\\
MZ-TH/05-09     \\
\end{flushright}

\vspace{0.6cm}

\begin{center}   

{\LARGE \sc Fractal Spacetime Structure in\\[5mm] Asymptotically Safe Gravity}

\vspace{2cm}
{\large O. Lauscher$^{(a)}$ and M. Reuter$^{(b)}$}\\

\vspace{1.7cm}
\noindent
{(a)\it Institute of Theoretical Physics, University of Leipzig\\
Augustusplatz 10-11, D-04109 Leipzig, Germany}\\[12pt]
{(b)\it Institute of Physics, University of Mainz\\
Staudingerweg 7, D-55099 Mainz, Germany}\\[6pt]
\end{center}   
 
\vspace*{1.6cm}

\begin{abstract}
Four-dimensional Quantum Einstein Gravity (QEG) is likely to be an 
asymptotically safe theory which is applicable at arbitrarily small distance
scales. On sub-Planckian distances it predicts that spacetime is a fractal with
an effective dimensionality of 2. The original argument leading to this result 
was based upon the anomalous dimension of Newton's constant. In the present
paper we demonstrate that also the spectral dimension equals 2 microscopically,
while it is equal to 4 on macroscopic scales. This result is an exact 
consequence of asymptotic safety and does not rely on any truncation. Contact 
is made with recent Monte Carlo simulations.
\end{abstract}
\end{titlepage}
%
\section{Introduction}
\renewcommand{\theequation}{1.\arabic{equation}}
\setcounter{equation}{0}
\label{3intro}
Soon after the detailed investigation of Quantum Einstein Gravity 
\cite{mr}-\cite{max} had begun and it had become clear that the theory is 
likely to be nonperturbatively renormalizable or ``asymptotically safe'' 
\cite{souma,oliver1,oliver2,oliver3} it was observed \cite{oliver1,oliver2}
that it predicts a fractal spacetime structure at sub-Planckian distances
whose effective dimensionality equals 2. On the technical side, a key tool in
the nonperturbative investigation of Quantum Einstein Gravity (QEG) was the
effective average action and its associated exact renormalization group (RG)
equation which had been developed in \cite{avact,ym} and was first applied to 
gravity in \cite{mr}. (For general reviews see \cite{avactrev},\cite{bagber}.)

In QEG, the effective average action $\Gamma_k[g_{\mu\nu}]$ defines an infinite
set of effective field theories, valid near a variable mass scale $k$ which
is introduced as an infrared (IR) cutoff and varies between $k=0$ and 
$k=\infty$ \cite{mr}. Intuitively speaking, the solution 
$\big<g_{\mu\nu}\big>_k$ of the scale dependent field equation 
\begin{eqnarray}
\label{fe}
\frac{\delta\Gamma_k}{\delta g_{\mu\nu}(x)}[\big<g\big>_k]=0
\end{eqnarray}
can be interpreted as the metric averaged over (Euclidean) spacetime volumes
of a linear extension $\ell$ which typically is of the order of $1/k$. Knowing
the scale dependence of $\Gamma_k$, i.e. the renormalization group trajectory
$k\mapsto\Gamma_k$, we can in principle follow the solution 
$\big<g_{\mu\nu}\big>_k$ from the ultraviolet $(k\rightarrow\infty)$ to the
infrared $(k\rightarrow 0)$.

It is an important feature of this approach that the infinitely many equations
of (\ref{fe}), one for each scale $k$, are valid {\it simultaneously}. They
all refer {\it to the same} physical system, the ``quantum spacetime'', but
describe its effective metric structure on different scales. An observer using
a ``microscope'' with a resolution $\approx k^{-1}$ will perceive the universe
to be a Riemannian manifold with metric $\big<g_{\mu\nu}\big>_k$. At every 
fixed $k$, $\big<g_{\mu\nu}\big>_k$ is a smooth classical metric. But since
the quantum spacetime is characterized by the infinity of equations (\ref{fe})
with $k=0,\cdots,\infty$ it can acquire very nonclassical and in particular
fractal features.

Let us describe more precisely what it means to ``average'' over Euclidean
spacetime
volumes. The quantity we can freely tune is the IR cutoff scale $k$, and the
``resolving power'' of the microscope, henceforth denoted $\ell$, is in
general a complicated function of $k$. (In flat space, $\ell\approx 1/k$.)
In order to understand the relationship between $\ell$ and $k$ we must recall
some steps from the construction of $\Gamma_k[g_{\mu\nu}]$ in ref. \cite{mr}.

The effective average action is obtained by introducing an IR cutoff into the
path-integral over all metrics, gauge fixed by means of a background gauge
fixing condition. Even without a cutoff the resulting effective action 
$\Gamma[g_{\mu\nu};\bar{g}_{\mu\nu}]$ depends on two metrics, the expectation
value of the quantum field, $g_{\mu\nu}$, and the background field 
$\bar{g}_{\mu\nu}$. This is a standard technique, and it is well known 
\cite{back} that the functional 
$\Gamma[g_{\mu\nu}]\equiv\Gamma[g_{\mu\nu};\bar{g}_{\mu\nu}=g_{\mu\nu}]$ 
obtained by equating the two
metrics can be used to generate the 1PI Green's functions of the theory.

The IR cutoff of the average action is implemented by first expressing the
functional integral over all metrics in terms of eigenmodes of $\bar{D}^2$,
the covariant Laplacian formed with the aid of the background metric 
$\bar{g}_{\mu\nu}$. Then a suppression term is introduced which damps the
contribution of all $-\bar{D}^2$-modes with eigenvalues smaller than $k^2$.
Following the usual steps \cite{avactrev} this leads to the scale dependent
functional $\Gamma_k[g_{\mu\nu};\bar{g}_{\mu\nu}]$, and the action with one
argument again obtains by equating the two metrics:
$\Gamma_k[g_{\mu\nu}]\equiv\Gamma_k[g_{\mu\nu};\bar{g}_{\mu\nu}=g_{\mu\nu}]$.
It is this action which appears in (\ref{fe}). Because of the identification
of the two metrics we see that it is basically the eigenmodes of
$\bar{D}^2=D^2$,
constructed from the argument of $\Gamma_k[g]$, which are cut off at $k^2$.
Since $\big<g_{\mu\nu}\big>_k$ is the corresponding stationary point, we can
say that the metric $\big<g_{\mu\nu}\big>_k$ applies to the situation where
only the quantum fluctuations of $-D^2(\big<g_{\mu\nu}\big>_k)$ with
eigenvalues larger than $k^2$ are integrated out. Therefore there is a 
complicated
interrelation between the metric and the scale at which it provides an
effective description: The covariant Laplacian which ultimately decides about
which modes are integrated out is constructed from the ``on shell'' 
configuration $\big<g_{\mu\nu}\big>_k$, so it is $k$-dependent by itself
already.

From these remarks it is clear now how to obtain the ``resolving power'' 
$\ell$ for a given $k$, at least in principle. We take the Laplacian 
$-D^2(\big<g_{\mu\nu}\big>_k)$, solve its eigenvalue problem, and then analyze 
in particular the properties of the eigenfunction(s) with eigenvalue $k^2$ 
(or near
$k^2$ in the case of a discrete spectrum). Loosely speaking, this eigenfunction
is the last one integrated out. As a consequence, the ``averaging'' scale is
crucially determined by  ``how fast'' this eigenfunction varies over 
spacetime. Let us assume, for instance, this eigenfunction is oscillatory with
a coordinate period $\Delta x$. Then, again by using
$\big<g_{\mu\nu}\big>_k$, we compute the physical proper length this
period corresponds to, and this is then what determines the resolution $\ell$.

In general the eigenfunction at $k^2$ will have a complicated $x$-dependence,
and therefore also the typical scales on which it varies are position 
dependent. Moreover, at a given point, the scale of significant variation will
be direction dependent (anisotropic). Therefore the resolving power
$\ell=\ell(k;x,n)$ is a complicated function in general, depending 
parametrically on points $(x^\mu)$ and directions $(n^\mu)$ on spacetime. It
is clear that these notions can be made precise only in concrete examples and
must be defined on a case by case basis.

We emphasize, however, that using the averaged metric itself to define the
scale it is averaged over is not a vicious circle, but rather is exactly as it
must be in a background independent approach to the quantization of gravity.

In a somewhat simplified form, the construction of a quantum spacetime within
QEG can be summarized as follows. We start from a fixed RG trajectory
$k\mapsto\Gamma_k$, derive its effective field equations at each $k$, and solve
them. The resulting quantum mechanical counterpart of a classical spacetime is 
specified by the infinity of Riemannian metrics $\{\big<g_{\mu\nu}\big>_k\big|
k=0,\cdots,\infty\}$. While the totality of these metrics contains all 
physical information, the parameter $k$ is only a book keeping device 
a priori. In a  second step, it can be given a physical interpretation by 
relating it
to the (proper) length scale of the averaging procedure: One constructs the
Laplacian $-D^2(\big<g_{\mu\nu}\big>_k)$, diagonalizes it, looks how rapidly 
its $k^2$-eigenfunction varies, and ``measures'' the length of typical 
variations with the metric $\big<g_{\mu\nu}\big>_k$ itself. By solving the
resulting $\ell=\ell(k)$ for $k=k(\ell)$ we can in principle reinterprete the
metric $\big<g_{\mu\nu}\big>_k$ as referring to a microscope with a known
position and direction dependent resolving power. The price we have to pay for
the background independence is that we cannot freely choose $\ell$ directly
but rather $k$ only.

The first difficult step in this construction program consists in finding the
RG trajectories. The running action $\Gamma_k[g_{\mu\nu}]$ is given by an
exact functional RG equation \cite{mr}. In practice it is usually solved on a
truncated theory space. In the Einstein-Hilbert truncation, for instance, 
$\Gamma_k$ is approximated by a functional of the form
\begin{eqnarray}
\label{3in2}
\Gamma_k[g]=\left(16\pi G_k\right)^{-1}\int d^4x\,\sqrt{g}\left\{
-R(g)+2\bar{\lambda}_k\right\}
\end{eqnarray}
involving a running Newton constant $G_k$ and cosmological constant 
$\bar{\lambda}_k$. Their $k$-dependence can be obtained from the RG equation
projected onto the truncation subspace. The $\beta$-functions for the 
dimensionless couplings $g_k\equiv k^2\,G_k$ and 
$\lambda_k\equiv\bar{\lambda}_k/k^2$ were first obtained in \cite{mr}. 
Remarkably, they turned out to possess a simultaneous zero at a non-Gaussian
fixed point (NGFP) $(g_*,\lambda_*)$ which has just the right properties needed
for the nonperturbative renormalizability of QEG along the lines of Weinberg's
\cite{wein} ideas on ``asymptotic safety'' \cite{souma}. It was argued 
\cite{oliver1,oliver2,oliver3} that the NGFP is likely to exist also in the
un-truncated, full theory and allows for the construction of a consistent and
predictive microscopic theory of quantum gravity valid at
arbitrarily small distances even.

One of the highly intriguing conclusions we reached in refs. 
\cite{oliver1,oliver2} was that the effective dimensionality of spacetime is
scale dependent. It equals 4 at macroscopic distances ($\ell\gg\ell_{\rm Pl}$)
but, near $\ell\approx\ell_{\rm Pl}$, it gets dynamically reduced to the value
2. For $\ell\ll\ell_{\rm Pl}$ spacetime is, in a precise sense \cite{oliver1},
a 2-dimensional fractal.

In ref. \cite{cosmo1} the specific form of the graviton propagator on this 
fractal was applied in a cosmological context. It was argued that it gives rise
to a Harrison-Zeldovich spectrum of primordial geometry fluctuations, perhaps
responsible for the CMBR spectrum observed today.

In refs. \cite{cosmo1}-\cite{mof} various types of ``RG improvements'' were
used to explore possible manifestations of the scale dependence of the
gravitational parameters.

Along a quite different line of investigation, considerable progress has been
made recently towards defining a quantum theory of gravity as the continuum
limit of a discrete model of statistical mechanics. Performing comprehensive
Monte Carlo simulations within the framework of causal (Lorentzian)
triangulated geometries \cite{amb}, Ambj\o rn, Jurkiewicz and Loll 
\cite{ajl1}-\cite{ajl34} collected strong evidence indicating that these models
can describe universes which are extended both in space and time and are
4-dimensional on large scales. In particular the above authors ``measured'' 
numerically the
spectral and Hausdorff dimensions of the spacetimes and their time slices,
respectively. Remarkably, they, too, find that the (spectral) dimension 
${\cal D}_{\rm s}$ of the spacetime reduces dynamically from 
${\cal D}_{\rm s}\approx 4$ at large distances to ${\cal D}_{\rm s}\approx 2$ 
on small length scales.

While until recently it has been difficult to compare the continuum theory to
the discrete causal triangulation approach, the new Monte Carlo results suggest
that they might be closely related, possibly representing the same 
``universality class''.

In our original argument \cite{oliver1} we determined the effective 
dimensionality of the fractal realized at sub-Planckian distances (in the
asymptotic scaling regime of the NGFP) from the anomalous dimension $\eta_N$
at the NGFP. A priori this definition of an effective dimensionality is 
different from the one used in the Monte Carlo simulations. It is the main
purpose of the present paper to apply the reasoning from \cite{oliver1,oliver2}
to the definition of the effective dimension which was employed by Ambj\o rn, 
Jurkiewicz and Loll \cite{ajl34}, namely the spectral dimension 
${\cal D}_{\rm s}$.

We shall demonstrate that asymptotically safe QEG does indeed predict
${\cal D}_{\rm s}=4$ at $\ell\gg\ell_{\rm Pl}$ and ${\cal D}_s=2$ for
$\ell\ll\ell_{\rm Pl}$. (The Planck length and mass are defined as 
$\ell_{\rm Pl}\equiv m_{\rm Pl}^{-1}\equiv G(k=0)^{1/2}$.)

As a preparation we review and extend the discussion of refs. 
\cite{oliver1,oliver2} in Section 2, and in Section 3 we compute the spectral
dimension of the QEG spacetimes.

\section{QEG spacetimes under the microscope}
\renewcommand{\theequation}{2.\arabic{equation}}
\setcounter{equation}{0}
\label{Sec2}
For simplicity we use the Einstein-Hilbert truncation to start with, and we
consider spacetimes with classical dimensionality $d=4$. The corresponding RG
trajectories were completely classified and determined numerically in 
\cite{frank1}. The physically relevant ones, for $k\rightarrow\infty$, all 
approach the NGFP at $(g_*,\lambda_*)$ so that the dimensionful quantities
run according to
\begin{eqnarray}
\label{asymrun}
G_k\approx g_*/k^2\;\;\;\;,\hspace{2cm}\bar{\lambda}_k\approx\lambda_*\,k^2
\end{eqnarray}
The behavior (\ref{asymrun}) is realized in the asymptotic scaling regime
$k\gg m_{\rm Pl}$. Near $k= m_{\rm Pl}$ the trajectories cross over towards
the Gaussian fixed point at $g=\lambda=0$, and then run towards negative,
vanishing, and positive values of $\lambda$, respectively.

Since in this paper we are interested only in the limiting cases of very small
and very large distances the following caricature of a RG trajectory will be
sufficient. We assume that $G_k$ and $\bar{\lambda}_k$ behave as in 
(\ref{asymrun}) for $k\gg m_{\rm Pl}$, and that they assume constant values
for $k\ll m_{\rm Pl}$. The precise interpolation between the two regimes could
be obtained numerically \cite{frank1} but will not be needed here.

The argument of ref. \cite{oliver2} concerning the fractal nature of the QEG
spacetimes was as follows. Within the Einstein-Hilbert truncation of theory
space, the effective field equations (\ref{fe}) happen to coincide with the 
ordinary Einstein equation, but with $G_k$ and $\bar{\lambda}_k$ replacing the
classical constants. Without matter,
\begin{eqnarray}
\label{einsteq}
R_{\mu\nu}(\big<g\big>_k)
&=&\bar{\lambda}_k\,\big<g_{\mu\nu}\big>_k
\end{eqnarray}
Since in absence of dimensionful constants of integration
$\bar{\lambda}_k$ is the only quantity in this equation which sets a 
scale, every solution to (\ref{einsteq}) has a typical radius of curvature 
$r_c(k)\propto 1/\sqrt{\bar{\lambda}_k}$. (For instance, the maximally 
symmetric $S^4$-solution has the radius $r_c=r=\sqrt{3/\bar{\lambda}_k}$.) 
If we want to explore the spacetime structure at a fixed length scale $\ell$ 
we should use the action $\Gamma_k[g_{\mu\nu}]$ at $k=1/\ell$ because with 
this functional a tree level analysis is sufficient to describe the essential 
physics at this scale, including the relevant quantum effects. Hence, when we 
observe the spacetime with a microscope of resolution $\ell$, we will see an 
average radius of curvature given by 
$r_c(\ell)\equiv r_c(k=1/\ell)$. Once $\ell$ is
smaller than the Planck length $\ell_{\rm Pl}\equiv m_{\rm Pl}^{-1}$
we are in the fixed point regime where $\bar{\lambda}_k\propto k^2$ so that 
$r_c(k)\propto 1/k$, or
\begin{eqnarray}
\label{radius}
r_c(\ell)\propto\ell
\end{eqnarray}
Thus, when we look at the structure of spacetime with a microscope of 
resolution $\ell\ll\ell_{\rm Pl}$, the average radius 
of curvature which we measure is proportional to the resolution 
itself. If we want to probe finer details and decrease $\ell$ we automatically
decrease $r_c$ and hence {\it in}crease the average curvature. Spacetime seems
to be more strongly curved at small distances than at larger ones. The 
scale-free relation (\ref{radius}) suggests that at distances below the Planck
length the QEG spacetime is a special kind of fractal with a self-similar 
structure. It has no intrinsic scale because in the fractal regime, i.e. when 
the RG trajectory is still close to the NGFP, the parameters which usually
set the scales of the gravitational interaction, $G$ and $\bar{\lambda}$, are 
not yet ``frozen out''. This happens only later on, somewhere half way between
the non-Gaussian and the Gaussian fixed point, at a scale of the order of 
$m_{\rm Pl}$.

Below this scale, $G_k $ and $\bar{\lambda}_k$ stop running and, as a result,
$r_c(k)$ becomes independent of $k$ so that $r_c(\ell)={\rm const}$ for 
$\ell\gg\ell_{\rm Pl}$. In this regime $\big<g_{\mu\nu}\big>_k$ is 
$k$-independent, indicating that the macroscopic spacetime is describable by a
single smooth, classical Riemannian manifold.

The above argument made essential use of the proportionality $\ell\propto 1/k$.
In the fixed point regime it follows trivially from the fact that there exist 
no relevant dimensionful parameters so that $1/k$ is the only length scale one 
can form. The algorithm for the determination of $\ell(k)$ described in the
Introduction yields the same answer.

It is easy to make the $k$-dependence of $\big<g_{\mu\nu}\big>_k$ explicit.
Picking an arbitrary reference scale $k_0$ we may rewrite (\ref{einsteq}) as
$[\bar{\lambda}_{k_0}/\bar{\lambda}_k]\,R^\mu_{\;\;\nu}(\big<g\big>_k)
=\bar{\lambda}_{k_0}\,\delta^\mu_\nu$. Since $R^\mu_{\;\;\nu}(c\,g)=c^{-1}\,
R^\mu_{\;\;\nu}(g)$ for any constant $c>0$, this relation implies that the
average metric scales as
\begin{eqnarray}
\label{metscale}
\big<g_{\mu\nu}(x)\big>_k&=&[\bar{\lambda}_{k_0}/\bar{\lambda}_k]\,
\big<g_{\mu\nu}(x)\big>_{k_0}
\end{eqnarray}
and its inverse according to
\begin{eqnarray}
\label{invmetscale}
\big<g^{\mu\nu}(x)\big>_k&=&[\bar{\lambda}_k/\bar{\lambda}_{k_0}]\,
\big<g^{\mu\nu}(x)\big>_{k_0}
\end{eqnarray}
These relations are valid provided the family of solutions considered exists
for all scales between $k_0$ and $k$, and $\bar{\lambda}_k$ has the
same sign always.

As we discussed in ref. \cite{oliver1} the QEG spacetime has an effective
dimensionality which is $k$-dependent and hence noninteger in general. Our
discussion in \cite{oliver1} was based upon the anomalous dimension $\eta_N$
of the operator $\int\sqrt{g}\,R$. It is defined as $\eta_N\equiv-k\,\partial_k
\ln Z_{Nk}$ where $Z_{Nk}\propto 1/G_k$ is the wavefunction renormalization
of the metric \cite{mr}. In a sense which we shall make more precise in a 
moment, the effective dimensionality of spacetime equals $4+\eta_N$. The RG
trajectories of the Einstein-Hilbert truncation (within its domain of validity)
have $\eta_N\approx 0$ for $k\rightarrow 0$\footnote{In the case of type IIIa
trajectories \cite{frank1,h3} the macroscopic $k$-value is still far above
$k_{\rm term}$, i.e. in the ``GR regime'' described in \cite{h3}.}
and $\eta_N\approx -2$ for 
$k\rightarrow\infty$, the smooth change by two units occuring near 
$k\approx m_{\rm Pl}$. As a consequence, the effective dimensionality is 4 for
$\ell\gg\ell_{\rm Pl}$ and 2 for $\ell\ll\ell_{\rm Pl}$.

In fact, the UV fixed point has an anomalous dimension $\eta\equiv\eta_N(g_*,
\lambda_*)=-2$. We can use this information in order to determine the 
momentum dependence of the dressed graviton propagator for momenta $p^2\gg
m_{\rm Pl}^2$. Expanding the $\Gamma_k$ of (\ref{3in2}) about flat space
and omitting the standard tensor structures we find the inverse propagator
$\widetilde{\cal G}_k(p)^{-1}\propto Z_N(k)\,p^2$. The conventional dressed 
propagator $\widetilde{\cal G}(p)$ contained in $\Gamma\equiv\Gamma_{k=0}$ 
obtains from the {\it exact} $\widetilde{\cal G}_k$ in the limit 
$k\rightarrow 0$.
For $p^2>k^2\gg m_{\rm Pl}^2$ the actual cutoff scale is the physical momentum
$p^2$ itself\footnote{See Section 1 of ref. \cite{h1} for a detailed discussion
of ``decoupling'' phenomena of this kind.}
so that the $k$-evolution of $\widetilde{\cal G}_k(p)$ stops at the threshold 
$k=\sqrt{p^2}$. Therefore
\begin{eqnarray}
\label{gp1}
\widetilde{\cal G}(p)^{-1}\propto\;Z_N\left(k=\sqrt{p^2}\right)\,p^2\propto\;
(p^2)^{1-\frac{\eta}{2}}
\end{eqnarray}
because $Z_N(k)\propto k^{-\eta}$ when $\eta\equiv-\partial_t\ln Z_N$ is 
(approximately) constant. In $d$ dimensions, and for $\eta\neq 2-d$, the
Fourier transform of $\widetilde{\cal G}(p)\propto 1/(p^2)^{1-\eta/2}$ yields 
the following propagator in position space:
\begin{eqnarray}
\label{gp2}
{\cal G}(x;y)\propto\;\frac{1}{\left|x-y\right|^{d-2+\eta}}\;.
\end{eqnarray}
This form of the propagator is well known from the theory of critical 
phenomena, for instance. (In the latter case it applies to large distances.)
Eq. (\ref{gp2}) is not valid directly at the NGFP. For $d=4$ and $\eta=-2$
the dressed propagator is $\widetilde{\cal G}(p)=1/p^4$ which has the following
representation in position space:
\begin{eqnarray}
\label{gp3}
{\cal G}(x;y)=-\frac{1}{8\pi^2}\,\ln\left(\mu\left|x-y\right|\right)\;.
\end{eqnarray}
Here $\mu$ is an arbitrary constant with the dimension of a mass. Obviously
(\ref{gp3}) has the same form as a $1/p^2$-propagator in 2 dimensions.

Slightly away from the NGFP, before other physical scales intervene, the 
propagator is of the familiar type (\ref{gp2}) which shows that the quantity 
$\eta_N$
has the standard interpretation of an anomalous dimension in the sense that
fluctuation effects modify the decay properties of ${\cal G}$ so as to 
correspond to a spacetime of effective dimensionality $4+\eta_N$.

Thus the properties of the RG trajectories imply a remarkable dimensional
reduction: Spacetime, probed by a ``graviton'' with $p^2\ll m_{\rm Pl}^2$ is
4-dimensional, but it appears to be 2-dimensional for a graviton with 
$p^2\gg m_{\rm Pl}^2$ \cite{oliver1}.

It is interesting to note that in $d$ classical dimensions, where the 
macroscopic spacetime is $d$-dimensional, the anomalous dimension at the
fixed point is $\eta=2-d$. Therefore, for any $d$, the dimensionality of the
fractal as implied by $\eta_N$ is $d+\eta=2$ \cite{oliver1,oliver2}.

\section{The spectral dimension}
\renewcommand{\theequation}{3.\arabic{equation}}
\setcounter{equation}{0}
\label{Sec3}
In this section we determine the spectral dimension ${\cal D}_{\rm s}$ of the
QEG spacetimes. This particular definition of a fractal dimension is borrowed
from the theory of diffusion processes on fractals \cite{avra} and is easily 
adapted to the
quantum gravity context \cite{nino,ajl34}. In particular it has been used in
the Monte Carlo studies mentioned in the Introduction.

Let us study the diffusion of a scalar test particle on a $d$-dimensional 
classical Euclidean manifold with a fixed smooth metric $g_{\mu\nu}(x)$.
The corresponding heat-kernel $K_g(x,x';T)$ giving the probability for the
particle to diffuse from $x'$ to $x$ during the fictitious diffusion time $T$
satisfies the heat equation
\begin{eqnarray}
\label{heateq}
\partial_T K_g(x,x';T)=\Delta_g K_g(x,x';T)
\end{eqnarray}
where $\Delta_g\equiv D^2$ denotes the scalar Laplacian: 
$\Delta_g\phi\equiv g^{-1/2}\,\partial_\mu(g^{1/2}\,g^{\mu\nu}\,\partial_\nu
\phi)$. The heat-kernel is a matrix element of the operator 
$\exp(T\,\Delta_g)$. In the random walk picture its trace per unit volume,
\begin{eqnarray}
\label{trace}
P_g(T)&\equiv& V^{-1}\int d^dx\,\sqrt{g(x)}\,K_g(x,x;T)\,\;\equiv\,\; 
V^{-1}\,{\rm Tr}\,
\exp(T\,\Delta_g)\;,
\end{eqnarray}
has the interpretation of an average return probability. (Here $V\equiv\int
d^dx\,\sqrt{g}$ denotes the total volume.) It is well known that $P_g$
possesses an asymptotic expansion (for $T\rightarrow 0$) of the form
$P_g(T)=(4\pi T)^{-d/2}\sum_{n=0}^\infty A_n\,T^n$. For an infinite flat
space, for instance, it reads $P_g(T)=(4\pi T)^{-d/2}$ for all $T$. Thus,
knowing the function $P_g$, one can recover the dimensionality of the target
manifold as the $T$-independent logarithmic derivative
\begin{eqnarray}
\label{dimform}
d=-2\frac{d\ln P_g(T)}{d\ln T}
\end{eqnarray}
This formula can also be used for curved spaces and spaces with finite volume
$V$ provided $T$ is not taken too large \cite{ajl34}.

In QEG where we functionally integrate over all metrics it is
natural to replace $P_g(T)$ by its expectation value. Symbolically,
\begin{eqnarray}
\label{pexpect}
P(T)\equiv\big<P_\gamma(T)\big>\equiv\int {\cal D}\gamma\,{\cal D}C\,{\cal D}
\bar{C}
\,\,P_\gamma(T)\,\;e^{-S_{\rm bare}[\gamma,C,\bar{C}]}
\end{eqnarray}
Here $\gamma_{\mu\nu}$ denotes the microscopic metric and $S_{\rm bare}$ is the
bare action with the gauge fixing terms and the pieces containing the ghosts
$C$ and $\bar{C}$ included. Note that (\ref{pexpect}) does not contain any IR
cutoff; it is the ordinary $(k=0)$ expectation value with all modes integrated
out. In QEG the functional $S_{\rm bare}$ is given by
the fixed point action. Given $P(T)$, the spectral dimension of the quantum
spacetime is defined in analogy with (\ref{dimform}):
\begin{eqnarray}
\label{specdim}
{\cal D}_{\rm s}=-2\frac{d\ln P(T)}{d\ln T}
\end{eqnarray}

Let us now evaluate the expectation value (\ref{pexpect}) using the average
action method. The fictitious diffusion process takes place on a ``manifold''
which, at every fixed scale, is described by a smooth Riemannian metric
$\big<g_{\mu\nu}\big>_k$. While the situation appears to be classical at fixed
$k$, nonclassical features emerge in the regime with nontrivial RG running
since there the metric depends on the scale at which the spacetime structure
is probed. 

The nonclassical features are encoded in the properties of the diffusion
operator. Denoting the covariant Laplacians corresponding to the 
metrics $\big<g_{\mu\nu}\big>_k$ and $\big<g_{\mu\nu}\big>_{k_0}$ by
$\Delta(k)$ and $\Delta(k_0)$, respectively, eqs. (\ref{metscale}) and 
(\ref{invmetscale}) imply that they are related by
\begin{eqnarray}
\label{opscale0}
\Delta(k)&=&[\bar{\lambda}_k/\bar{\lambda}_{k_0}]\,\Delta(k_0)
\end{eqnarray}
When $k,k_0\gg m_{\rm Pl}$ we have, for example, 
\begin{eqnarray}
\label{opscale}
\Delta(k)&=&(k/k_0)^2\,\Delta(k_0)
\end{eqnarray}

Recalling that the average action $\Gamma_k$ defines an effective field 
theory at the scale $k$ we have that $\big<{\cal O}(\gamma_{\mu\nu})\big>
\approx{\cal O}(\big<g_{\mu\nu}\big>_k)$ if the operator ${\cal O}$ involves
typical covariant momenta of the order $k$ and $\big<g_{\mu\nu}\big>_k$ solves
eq. (\ref{fe}). In the following we exploit this relationship for the
RHS of the diffusion equation, ${\cal O}=\Delta_\gamma\,K_\gamma(x,x';T)$.
It is crucial here to correctly identify the relevant scale $k$.

If the diffusion process  involves (approximately) only a small interval of 
scales near $k$ over which $\bar{\lambda}_k$ does not change much the 
corresponding heat equation contains the $\Delta(k)$ for this specific, fixed
value of $k$:
\begin{eqnarray}
\label{heateqk}
\partial_T K(x,x';T)&=&\Delta(k) K(x,x';T)
\end{eqnarray}
Denoting the eigenvalues of $-\Delta(k_0)$ by ${\cal E}_n$ and the 
corresponding eigenfunctions by $\phi_n$, this equation is solved by
\begin{eqnarray}
\label{kernexp}
K(x,x';T)&=&\sum\limits_n\phi_n(x)\,\phi_n(x')\,\exp\bigg(-F(k^2)\,{\cal E}_n
\,T\bigg)
\end{eqnarray}
Here we introduced the convenient notation $F(k^2)\equiv\bar{\lambda}_k/
\bar{\lambda}_{k_0}$. Knowing this propagation kernel we can time-evolve any
initial probability distribution $p(x;0)$ according to
$p(x;T)=\int d^4x'\,\sqrt{g_0(x')}\,K(x,x';T)\,p(x';0)$ with $g_0$ the 
determinant of $\big<g_{\mu\nu}\big>_{k_0}$. If the initial distribution has 
an eigenfunction expansion of the form $p(x;0)=\sum_n C_n\,\phi_n(x)$ we
obtain
\begin{eqnarray}
\label{probexp}
p(x;T)&=&\sum_n C_n\,\phi_n(x)\,\exp\bigg(-F(k^2)\,{\cal E}_n\,T\bigg)
\end{eqnarray}
If the $C_n$'s are significantly different from zero only for a single
eigenvalue ${\cal E}_N$, we are dealing with a single-scale problem. In the
usual spirit of effective field theories we would then identify 
$k^2={\cal E}_N$ as the relevant scale at which the running couplings are to
be evaluated.

However, in general the $C_n$'s are different from zero over a wide range of
eigenvalues. In this case we face a multiscale problem where different modes
$\phi_n$ probe the spacetime on different length scales.

If $\Delta(k_0)$ corresponds to flat space, say, the eigenfunctions $\phi_n
\equiv\phi_p$ are plane waves with momentum $p^\mu$, and they resolve
structures on a length scale $\ell$ of order $1/|p|$. Hence, in terms of the
eigenvalue ${\cal E}_n\equiv{\cal E}_p=p^2$ the resolution is $\ell\approx
1/\sqrt{{\cal E}_n}$. This suggests that when the manifold is probed by a
mode with eigenvalue ${\cal E}_n$ it ``sees'' the metric 
$\big<g_{\mu\nu}\big>_k$ for the scale $k=\sqrt{{\cal E}_n}$. Actually the
identification $k=\sqrt{{\cal E}_n}$ is correct also for curved space since,
in the construction of $\Gamma_k$, the parameter $k$ is introduced precisely
as a cutoff in the spectrum of the covariant Laplacian. 

Therefore we conclude that under the spectral sum of (\ref{probexp}) we must
use the scale $k^2={\cal E}_n$ which depends explicitly on the resolving power
of the
corresponding mode. Likewise, in eq. (\ref{kernexp}), $F(k^2)$ is to be
interpreted as $F({\cal E}_n)$. Thus we obtain the traced propagation kernel
\begin{eqnarray}
\label{trpropk}
P(T)&=&V^{-1}\;\sum_n \exp\bigg[-F({\cal E}_n)\,{\cal E}_n\,T\bigg]\nonumber\\
&=&V^{-1}\;{\rm Tr}\,\exp\bigg[F\Big(-\Delta(k_0)\Big)\,\Delta(k_0)\,T\bigg]
\end{eqnarray}

It is convenient to choose $k_0$ as a macroscopic scale in a regime where there
are no strong RG effects any more.

Furthermore, let us assume for a moment that at $k_0$ the cosmological
constant is tiny, $\bar{\lambda}_{k_0}\approx 0$, so that $\big<g_{\mu\nu}
\big>_{k_0}$ is an approximately flat metric. In this case the trace in eq.
(\ref{trpropk}) is easily evaluated in a plane wave basis:
\begin{eqnarray}
\label{trplane}
P(T)=\int\frac{d^4p}{(2\pi)^4}\,\exp\left[-p^2\,F(p^2)\,T\right]
\end{eqnarray}
The $T$-dependence of (\ref{trplane}) determines the fractal dimensionality of
spacetime via (\ref{specdim}). In the limits $T\rightarrow\infty$ and 
$T\rightarrow 0$ where the random walks probe very large and small distances,
respectively, we obtain the dimensionalities corresponding to the largest
and smallest length scales possible. The limits $T\rightarrow\infty$ and 
$T\rightarrow 0$ of $P(T)$ are determined by the behavior of $F(p^2)\equiv
\bar{\lambda}(k=\sqrt{p^2})/\bar{\lambda}_{k_0}$ for $p^2\rightarrow 0$ and
$p^2\rightarrow\infty$, respectively.

For a RG trajectory where the renormalization effects stop below some threshold
we have $F(p^2\rightarrow 0)=1$. In this case (\ref{trplane}) yields
$P(T)\propto 1/T^2$, and we conclude that the macroscopic spectral dimension
is ${\cal D}_{\rm s}=4$.

In the fixed point regime we have $\bar{\lambda}_k\propto k^2$, and therefore
$F(p^2)\propto p^2$. As a result, the exponent in (\ref{trplane}) is 
proportional to $p^4$ now. This implies the $T\rightarrow 0-$behavior
$P(T)\propto 1/T$. It corresponds to the spectral dimension 
${\cal D}_{\rm s}=2$.

This result holds for all RG trajectories since only the fixed point 
properties were used. In particular it is independent of $\bar{\lambda}_{k_0}$
on macroscopic scales. In fact, the above assumption that $\big<g_{\mu\nu}
\big>_{k_0}$ is flat was not necessary for obtaining ${\cal D}_{\rm s}=2$.
This follows from the fact that even for a curved metric the spectral sum
(\ref{trpropk}) can be represented by an Euler-Mac Laurin series which always
implies (\ref{trplane}) as the leading term for $T\rightarrow 0$.

Thus we may conclude that on very small and very large length scales the
spectral dimensions of the QEG spacetimes are
\begin{eqnarray}
\label{qegspecdims}
{\cal D}_{\rm s}(T\rightarrow\infty)&=&4\nonumber\\
{\cal D}_{\rm s}(T\rightarrow 0)&=&2
\end{eqnarray}

The dimensionality of the fractal realized at sub-Planckian distances is
found to be 2 again. It is by no means trivial that ${\cal D}_{\rm s}$ 
coincides with the value of $4+\eta$.
While the replacement of the classical $p^2\,F(p^2)=p^2$ by $p^2\,F(p^2)
\propto p^4$ is reminiscent of the graviton propagator argument of Section 2, 
it is important to emphasize that the value of $4+\eta$ is entirely determined
by the running of $G_k$, while the spectral dimension was derived from the
$k$-dependence of the cosmological constant.

In fact, it is remarkable that the equality of $4+\eta$ and ${\cal D}_{\rm s}$
is a special feature of 4 classical dimensions. Generalizing for $d$
classical dimensions, the fixed point running of Newton's constant becomes
$G_k\propto k^{2-d}$ with a dimension-dependent exponent, while 
$\bar{\lambda}_k\propto k^2$ continues to have a quadratic $k$-dependence. As
a result, the $\widetilde{{\cal G}}(k)$ of eq. (\ref{gp1}) is proportional to 
$1/p^d$ in general so that, for any $d$, the 2-dimensional looking graviton
propagator (\ref{gp3}) is obtained. (This is equivalent to saying that
$\eta=2-d$, or $d+\eta=2$, for arbitrary $d$.)

On the other hand, the impact of the RG effects on the diffusion process is to
replace the operator $\Delta$ by $\Delta^2$, for any $d$, since the 
cosmological constant always runs quadratically. Hence, in the fixed point
regime, eq. (\ref{trplane}) becomes
\begin{eqnarray}
\label{trplanefp}
P(T)\propto\int d^dp\,\exp\left[-p^4\,T\right]\propto T^{-\frac{d}{4}}
\end{eqnarray}
This $T$-dependence implies the spectral dimension
\begin{eqnarray}
\label{qegspecdims2}
{\cal D}_{\rm s}(d)&=&\frac{d}{2}
\end{eqnarray}
This value coincides with $d+\eta$ if, and only if, $d=4$. It is an intriguing
speculation that this could have something to do with the observed macroscopic
dimensionality of spacetime.

Up to this point, to be as concrete as possible, we formulated our argument
within the Einstein-Hilbert truncation. To complete the discussion we 
show that the exact (un-truncated) theory if it has a NGFP implies the 
dynamical dimensional
reduction from 4 to 2 dimensions (in $d=4$) in exactly the same way as the 
truncated one.

The complete effective average action has the structure 
$\Gamma_k[g_{\mu\nu}]=\sum_n\bar{\rm g}_n(k)\,I_n[g_{\mu\nu}]$ with infinitely
many running couplings $\bar{\rm g}_n(k)$ and diffeomorphism invariant 
functionals $I_n$. If $\bar{\rm g}_n(k)$ has the canonical dimension $d_n$ the
corresponding dimensionless couplings are ${\rm g}_n(k)\equiv k^{-d_n}\,
\bar{\rm g}_n(k)$ and we have
\begin{eqnarray}
\label{fullact}
\Gamma_k[g_{\mu\nu}]&=&\sum_n{\rm g}_n(k)\,k^{d_n}\,I_n[g_{\mu\nu}]\;\,=\;\,
\sum_n{\rm g}_n(k)\,I_n[k^2\,g_{\mu\nu}]
\end{eqnarray}
In the second equality we used that $I_n[c^2\,g_{\mu\nu}]=c^{d_n}\,
I_n[g_{\mu\nu}]$ for any $c>0$ since $I_n$ has dimension $-d_n$.

If the theory is asymptotically safe at the exact level, all
${\rm g}_n(k)$ approach constant values ${\rm g}_{n*}$ for 
$k\rightarrow\infty$:
\begin{eqnarray}
\label{fixpact}
\Gamma_{k\rightarrow\infty}[g_{\mu\nu}]&=&\sum_n{\rm g}_{n*}\,
I_n[k^2\,g_{\mu\nu}]
\end{eqnarray}
Obviously this functional depends on $k^2$ and $g_{\mu\nu}$ only via the
combination $k^2\,g_{\mu\nu}$. Therefore the solutions of the corresponding
field equation, $\big<g_{\mu\nu}\big>_k$, scale proportional to $k^{-2}$. Hence
$\Delta(k)\propto k^2$ in the fixed point regime, and this is exactly the
scaling behavior (\ref{opscale}) our above derivation of ${\cal D}_{\rm s}
(T\rightarrow 0)=2$ was based upon.

This completes the demonstration that if a NGFP does exist in the full theory,
its exact spacetimes are fractals with ${\cal D}_{\rm s}=2$ on sub-Planckian
distances.

At this point it is tempting to compare the result (\ref{qegspecdims}) to the
spectral dimensions of the spacetime which were recently obtained by Monte
Carlo simulations of the causal dynamical triangulation model \cite{ajl34}:
\begin{eqnarray}
\label{mcspecdims}
{\cal D}_{\rm s}(T\rightarrow\infty)&=&4.02\pm 0.1\nonumber\\
{\cal D}_{\rm s}(T\rightarrow 0)&=&1.80\pm 0.25
\end{eqnarray}
These figures, too, suggest that the long-distance and short-distance spectral
dimension should be 4 and 2, respectively.

The dimensional reduction from 4 to 2 dimensions is a highly nontrivial
dynamical phenomenon which seems to occur in both QEG and the discrete
triangulation model. We find it quite remarkable that the discrete
and the continuum approach lead to essentially identical conclusions in this
respect. We consider this agreement a first hint indicating that (at least in
4 dimensions) the discrete model and QEG in the average action formulation
might describe the same physics. But clearly much more work is needed in
order to understand how the two approaches are related precisely.

\section{Summary}
\renewcommand{\theequation}{4.\arabic{equation}}
\setcounter{equation}{0}
\label{conclusio}
The general picture of the spacetime structure in QEG which has emerged so far
is as follows. At sub-Planckian distances spacetime is a fractal of
dimensionality ${\cal D}_{\rm s}=4+\eta=2$. It can be thought of as a
self-similar hierarchy of superimposed Riemannian manifolds of any curvature.
As one considers larger length scales where the RG running of the
gravitational parameters comes to a halt, the ``ripples'' in the spacetime
gradually disappear and the structure of a classical 4-dimensional manifold is
recovered.

Within the Einstein-Hilbert approximation of QEG there are two natural ways of
defining an effective dimensionality of the fractal spacetime. We can either
define it as $4+\eta_N$ as derived from the running of Newton's constant,
or we use the spectral dimension implied by the $k$-dependence of the
cosmological constant. We have seen that both definitions lead to identical
results on very large and very small distances. We also showed that the
microscopic dimensionality ${\cal D}_{\rm s}=2$ is a rather direct and exact
consequence of asymptotic safety which does not rely on any approximation or
truncation. It is therefore not unlikely
that the mechanism of a dynamical dimensional reduction from 4 to 2 dimensions
which occurs in QEG is the same phenomenon as the dimensional reduction
observed in the Monte Carlo studies of causal dynamical triangulations.\\[24pt]
Acknowledgments: We would like to thank J. Schwindt for helpful discussions. 

\end{document}